\documentclass{PoS}

\title{Anomalies and discrete chiral symmetries}

\ShortTitle{Anomalies and chiral symmetry}

\author{\speaker{Michael Creutz} \thanks{ I am grateful to the
 Alexander von Humboldt Foundation for support for my ongoing visit to
 the University of Mainz.  This manuscript has been authored under
 contract number DE-AC02-98CH10886 with the U.S.~Department of Energy.
 Accordingly, the U.S. Government retains a non-exclusive,
 royalty-free license to publish or reproduce the published form of
 this contribution, or allow others to do so, for U.S.~Government
 purposes.  }\\ Brookhaven National Laboratory and Universit\"at
 Mainz\\ E-mail: \email{creutz@bnl.gov}}

\abstract{The quantum anomaly that breaks the U(1) axial symmetry of
massless multi-flavored QCD leaves behind a discrete flavor-singlet
chiral invariance. With massive quarks, this residual symmetry has a
close connection with the strong CP-violating parameter theta. One
result is that if the lightest quarks are degenerate, then a first
order transition will occur when theta passes through pi.  The
resulting framework helps clarify when the rooting prescription for
extrapolating in the number of flavors is valid. }

\FullConference{International Workshop on QCD Green's Functions, 
Confinement and Phenomenology\\
         September 7-11 , 2009\\
         ECT Trento, Italy}

\begin{document}

\def \li {\par\hskip .2in {$\bullet$}\hskip .1 in }
\def \lli {\par            {$\bullet$}\hskip .1 in }

\def \hi {\medskip}
\def \topic #1{\par\centerline {#1}}

\def\fitframe #1#2#3{\vbox{\hrule height#1pt
 \hbox{\vrule width#1pt\kern #2pt
 \vbox{\kern #2pt\hbox{#3}\kern #2pt}
 \kern #2pt\vrule width#1pt}
 \hrule height0pt depth#1pt}}

\long \def \blockcomment #1\endcomment{}
\def \nextslide{\medskip}

\section{Introduction}

As is well known to this community, chiral symmetry plays a crucial
role in our understanding of QCD.  But, of course, chiral symmetry is
broken, and there are three sources of this breaking.  First is the
spontaneous breaking giving rise to an expectation value for the
chiral condensate; {\it i.e.}  $\langle\overline\psi\psi\rangle\ne 0$.
This is invoked to explain the lightness of pions relative to other
hadrons.  Second, we have the implicit breaking of the flavor-singlet
axial $U(1)$ symmetry by the anomaly.  This explains why the
$\eta^\prime$ meson is not so light in comparison with the pions.  And
finally we have the explicit breaking of chiral symmetry by the quark
masses.  This means that the pions, while light, are not exactly
massless.  In this talk I will discuss some of the rather rich physics
that arises from the interplay of these three effects.

The breaking of the classical $U(1)$ axial symmetry is tied to the
possibility of introducing into massive QCD a CP violating parameter,
usually called $\Theta$.  For a recent review of this quantity, see
Ref.~\cite{Vicari:2008jw}. One of my goals here is to provide an
intuitive and qualitative picture of the $\Theta$ parameter in meson
physics.  This picture has evolved over many years.  The possibility
of the spontaneous CP violation occurring at $\Theta=\pi$ is tied to
what is known as Dashen's phenomenon \cite{dashen}, first noted even
before the days of QCD.  In the mid 1970's, 't Hooft \cite{'tHooft:fv}
elucidated the underlying connection between the chiral anomaly and
the topology of gauge fields.  Later Witten \cite{Witten:1980sp} used
large gauge group ideas to discuss the behavior at $\Theta=\pi$ in
terms of effective Lagrangians.  Ref.~\cite{oldeffective} lists a few
of the early studies of the effects of $\Theta$ on effective
Lagrangians.  The topic continues to appear in various contexts; for
example, Ref.~\cite{Boer:2008ct} contains a different approach to
understanding the transition at $\Theta=\pi$ in the framework of the
two-flavor Nambu Jona-Lasinio model.

I became interested in these issues while trying to understand the
difficulties with formulating chiral symmetry on the lattice.  Much of
the picture presented here is implicit in my 1995 paper on quark
masses \cite{Creutz:1995wf}.  Since then the topic has become highly
controversial, with the realization of ambiguities precluding a
vanishing up quark mass as a solution to the strong CP problem
\cite{Creutz:2003xc} and the appearance of an inconsistency with one
of the popular algorithms in lattice gauge theory
\cite{Creutz:2008nk}.  Despite the controversies, both results are
immediate consequences of the interplay of the anomaly and chiral
symmetry.  The fact that these issues remain so disputed drives me to
return to them here.  Portions of this discussion appear in more
detail in Ref. \cite{Creutz:2009kx}.

A crucial issue is that the axial anomaly in $N_f$ flavor massless QCD
leaves behind a residual $Z_{N_f}$ flavor-singlet chiral symmetry.
This is closely tied to gauge-field topology and the QCD theta
parameter.  As a consequence I will show that, with degenerate quarks
carrying a small non-zero mass, there must appear a first order
transition at $\Theta=\pi$.  For two flavors this transition  studied
in Refs.~\cite{Creutz:1995wf,Smilga:1998dh,Tytgat:1999yx}.
This result in turn has several further
consequences.  First, the sign of the quark mass is relevant for an
odd number of flavors.  This is not seen in perturbation theory,
giving a simple example where perturbation theory does not provide a
complete description of a field theory.  Second, going down to one
flavor, chiral symmetry no longer provides an additive protection for
a small fermion mass.  And third, the nontrivial dependence on the
number of flavors can in some cases invalidate the rooting
prescription often used to extrapolate between different flavor
contents in lattice simulations.  It is the latter two points which
have been extremely controversial.

\section{Assumptions}

For the purposes of this talk I make a few minimal assumptions.
Considering QCD with $N_f$ light quarks, I assume this field theory
exists and confines in the usual way.  I assume that spontaneous
chiral symmetry breaking occurs in the massless theory with
$\langle\overline\psi\psi\rangle\ne0$.  When masses are considered, I
consider that the usual chiral perturbation theory in momenta and
masses makes sense.  I also assume that the anomaly gives the
$\eta^\prime$ a mass even when the quark masses vanish.  I further
consider $N_f$ small enough to avoid any possible conformal phases.

I frame the discussion in continuum language, but I imagine some
non-perturbative regulator is in place to control divergences.  Of
course for me this would be the lattice, but I need not be more
specific here.  I assume this regulator has brought us close to the
continuum theory, {\it i.e.} any momentum space cutoff should be much
larger than $\Lambda_{QCD}$, the natural scale of the strong
interactions.  For a lattice approach, the lattice spacing $a$ is
considered as much smaller than $1/\Lambda_{QCD}$.

I consider the effective potential $V$ for various meson fields.  This
represents the resulting vacuum energy density for a given field
expectation.  Such can be derived formally via a Legendre
transformation in the standard way.  Here I will ignore convexity
issues associated with the phase separation that will occur when a
field is constrained to be in a naively concave region.  A more
precise treatment would be in terms of the phase transitions that
occur with global minimum changes.  Instead I proceed with the
generally familiar language of symmetry breaking in terms of multiple
minima in the effective potential.  For simplicity I concentrate on
$N_f$ degenerate quarks.  To start I will also take $N_f$ even; this
is because of some interesting subtleties with an odd number of
flavors that I will get to later in the talk.

I will be considering a variety of composite fields.  Because these
are generally singular products of fields at the same space time point,
I assume that our unspecified regulator has some way of handling
this.  The particular fields I will work with are
\begin{equation}
\matrix{
\sigma\sim\overline\psi \psi\cr
\pi_\alpha\sim i\overline\psi \lambda_\alpha \gamma_5 \psi\cr
\eta^\prime \sim i\overline\psi\gamma_5\psi.\cr}
\end{equation}
Here $\lambda_\alpha$ represents the generalized Gell-Mann matrices
generating the flavor group $SU(N_f)$.

\section{Spontaneous chiral symmetry breaking}

Spontaneous breaking of chiral symmetry is a crucial part of our
understanding of the strong interactions.  It is usually discussed in
terms of a double well structure for the effective potential
considered as a function of the field $\sigma\sim\overline\psi \psi$.
The vacuum selects one of these minima giving an expectation value to
the sigma field, $\langle\sigma\rangle=v\ne 0$.  When the mass
vanishes it is a convention whether one takes the positive or the
negative minimum.

\begin{figure*}
\centering
\includegraphics[width=2in]{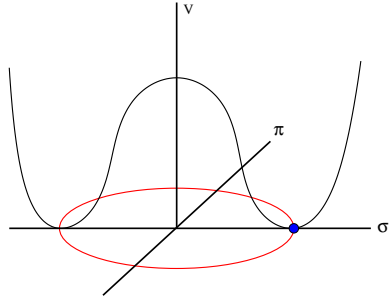}
\caption{\label{v0} Spontaneous chiral symmetry breaking is
represented by a double well effective potential with the vacuum
settling into a non-trivial minimum.  Chiral symmetry is broken by the
selection of a specific value for the quark condensate.  The flavor
non-singlet pseudo-scalar mesons are Goldstone bosons corresponding to
flat directions in the effective potential.  }
\end{figure*}

With multiple flavors the vacuum is continuously degenerate, with the
non-singlet pseudo-scalars being Goldstone bosons.  This is associated
with a symmetry under flavored chiral rotations of the quark fields
\begin{equation}\matrix{
\psi\rightarrow e^{i\phi\gamma_5\lambda^\alpha/2}\psi\cr
\overline\psi\rightarrow \overline\psi e^{i\phi\gamma_5\lambda^\alpha/2}.\cr
}
\end{equation}
There is one such symmetry for each generator of the flavor group
$SU(N_f)$.  For example, with two flavors this symmetry mixes the
$\sigma$ and $\pi$ fields 
\begin{equation}
\matrix{
&\sigma\rightarrow \cos(\phi)\sigma+ \sin(\phi)\pi^\alpha,\cr
&\pi^\alpha\rightarrow \cos(\phi)\pi^\alpha- \sin(\phi)\sigma.}
\end{equation}
The minimum of the potential has $N_f^2-1$ ``flat'' directions.  This
standard scenario is illustrated in Fig.~\ref{v0}.

\begin{figure*}
\centering
\includegraphics[width=2in]{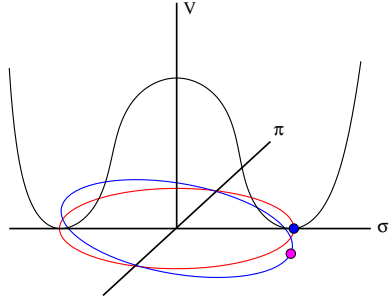}
\caption{\label{v3} A small quark mass tilts the effective potential,
selecting one direction for the true vacuum and giving the Goldstone
bosons a mass.  }
\end{figure*}

If we now consider a small quark mass, this will select one vacuum as
unique.  Physically, a mass term represented by $V \rightarrow
V-m\sigma$ tilts the effective potential downward in a specific
direction, as illustrated in Fig.~\ref{v3}.  In the process the
Goldstone bosons acquire a mass proportional to the square root of the
quark mass.

\section{The chiral anomaly}

It is the chiral anomaly that gives the flavor-singlet $\eta^\prime$ a
mass even if the quark mass vanishes.  This mass is of order the scale
of the strong interactions, $m_{\eta^\prime} = O(\Lambda_{QCD})$, and
does not go to zero when the quark mass does.  In terms of the
effective potential, $V(\sigma,\eta^\prime)$ is not symmetric under
the rotation
\begin{equation}
\matrix{
&\sigma\rightarrow \sigma \cos(\phi)+\eta^\prime \sin(\phi)\cr
&\eta^\prime\rightarrow 
\eta^\prime \cos(\phi)
-\sigma \sin(\phi).
\cr
}
\end{equation}
If we expand the effective potential near the vacuum state $\sigma\sim
v$ and $\eta^\prime \sim 0$ we should expect a form like
\begin{equation}
V(\sigma,\eta^\prime)= m_\sigma^2 (\sigma-v)^2 
+ m_{\eta^\prime}^2 {\eta^\prime}^2 
+O((\sigma-v)^3,{\eta^\prime}^4)
\end{equation}
with both masses being of order $\Lambda_{QCD}$.

In quark language, the above rotation mixing the $\sigma$ and
$\eta^\prime$ fields is associated with the classical symmetry of the
naive action under
\begin{equation}
\label{anom}
\matrix{
&\psi \rightarrow e^{i\phi\gamma_5/2}\psi\cr
&\overline\psi \rightarrow \overline\psi e^{i\phi\gamma_5/2}.\cr
}
\end{equation}
This symmetry is ``anomalous'' in the sense that it must be broken by
any valid regulator.  The $\eta^\prime$ mass is a remnant of this
breaking that survives as the regulator is removed.

Fujikawa \cite{Fujikawa:1979ay} has presented a rather elegant way to
see how the anomaly arises.  The above variable change
alters the fermion measure
\begin{equation}
d\psi\rightarrow |e^{-i\phi\gamma_5/2}| d\psi
=e^{-i\phi {\rm Tr}\gamma_5/2} d\psi.
\end{equation}
Now naively $\gamma_5$ is a traceless matrix, and one might conclude
that this change in the measure is harmless.  Fujikawa pointed out
that this does not apply in the regulated theory.  For example one
might define the trace of $\gamma_5$ as
\begin{equation}
{\lim_{\Lambda\rightarrow \infty}}
{\rm Tr} \left(\gamma_5 e^{-D^2/\Lambda^2}\right)\ne 0
\end{equation}
where $D$ is the kinetic part of the Dirac action $\overline\psi
(D+m)\psi$.  In the usual continuum analysis this satisfies
$D^\dagger=-D$ and anti-commutes with gamma five, $[D,\gamma_5]_+=0$.
Thus motivated, we can use the eigenstates of $D$ 
\begin{equation}
 D|\psi_i\rangle=\lambda_i|\psi_i\rangle
\end{equation}
to define the trace
\begin{equation}
{\rm Tr}\gamma_5=\sum_i  \langle\psi_i|\gamma_5|\psi_i\rangle.
\end{equation}

At this point we bring in the index theorem; this states that if the
gauge field has non-trivial winding $\nu$, $D$ will have at least
$\nu$ zero modes $D|\psi_i\rangle=0$.  These modes are chiral:
$\gamma_5|\psi_i\rangle=\pm |\psi_i\rangle$ and the counting is such
that $\nu=n_+-n_-$.  Thus the zero modes contribute $\nu$ to the trace
of $\gamma_5$.

Now the non-zero eigenmodes all occur in complex conjugate pairs.  If
we have $D|\psi\rangle=\lambda|\psi\rangle$, then $ D\gamma_5
|\psi\rangle=-\lambda\gamma_5 |\psi\rangle
=\lambda^*\gamma_5|\psi\rangle$.  As $D$ is anti-hermitian,
$|\psi\rangle$ and $|\gamma_5\psi\rangle$ are orthogonal whenever
$\lambda\ne 0$.  As a consequence, the space spanned by $|\psi\rangle$
and $|\gamma_5 \psi\rangle$ gives no contribution to ${\rm
Tr}\gamma_5$.  We are led to the remarkable conclusion that only the
zero modes count in calculating the above trace.  Thus we have
\begin{equation}
{\rm Tr}\gamma_5=\sum_i \langle \psi_i |\gamma_5|\psi_i\rangle=\nu,
\end{equation}
which does not vanish when the topology is non-trivial.

So where did the opposite chirality states go?  In continuum language,
they are ``lost at infinity'' in the sense that they have been driven
``above the cutoff.''  On the lattice there are no infinities; so
things are a bit more subtle.  With the overlap operator
\cite{Ginsparg:1981bj}, all eigenvalues lie on a circle in the complex
plane, and corresponding to every zero mode is a corresponding mode of
opposite chirality on the opposite side of this circle.  This
technique brings in a modified chiral matrix through the relation
$D\gamma_5=-\hat\gamma_5 D$ and the winding appears via ${\rm Tr}\
\hat\gamma_5=2\nu$.  With Wilson fermions \cite{Wilson:1975id} the low
lying approximate zero modes are compensated by additional real
eigenvalues in the doubler region.

Note that this discussion involves both short and long distances.  The
zero modes associated with topology are compensated by additional
modes lost at the cutoff.  This means that it can be dangerous to
assume that one can ignore instanton physics by going to short
distances.  Furthermore it becomes impossible to uniquely separate
perturbative and non-perturbative effects; as one changes, say, the
scale of the cutoff, small instantons can ``fall through the
lattice.''  In general this issue is scheme dependent.

So we conclude that under the transformation of Eq.~(\ref{anom}), the
regulated fermion measure changes by $e^{-i\phi {\rm
Tr}\gamma_5}=e^{-i\phi\nu}$.  This factor changes the weighting of
gauge configurations with non-zero winding.  Note that this introduces
a sign problem for Monte Carlo, but that is not the topic under
discussion here.

To end this section, note that the angle $\phi$ I have used here is
the conventional $\Theta/N_f$.  This is since I have given each flavor
a common phase.  Each contributes equally, and the full trace
including flavor space is ${\rm Tr}\gamma_5=N_f \nu$.

\section{A $Z_{N_f}$ symmetry}

\begin{figure*}
\centering
\includegraphics[width=2in]{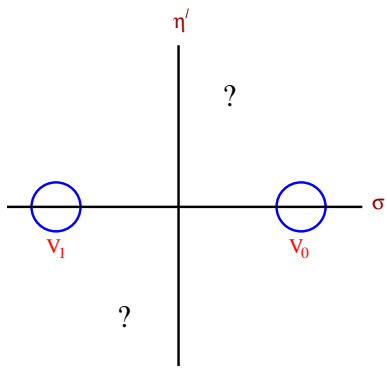}
\caption{\label{potential0} The circles in this figure inclose the two
minima in the $\sigma,\eta^\prime$ plane located at $\sigma=\pm v$ and
$\eta^\prime=0$.  Can we find any other minima?  }
\end{figure*}

I now return to the earlier effective-potential language.  I have
argued that, because of spontaneous chiral symmetry breaking, there
are at least two minima in the $\sigma,\eta^\prime$ plane, located as
sketched in Fig.~\ref{potential0}.  Do we know anything about the
potential elsewhere in the $\sigma,\eta^\prime$ plane?  Remarkably the
answer is yes; there are actually $N_f$ physically equivalent minima
in this plane.

\begin{figure*}
\centering
\includegraphics[width=2in]{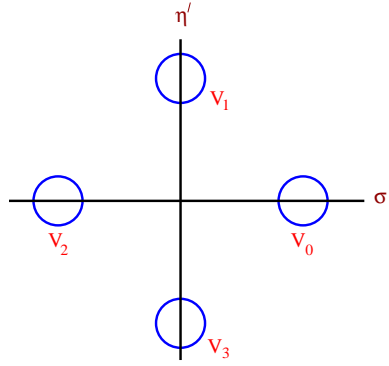}
\caption{\label{potential1} For four flavors, the effective potential
has four equivalent minima, marked here with circles, in the
$\sigma,\eta^\prime$ plane.  This generalizes to $N_f$ minima with
$N_f$ flavors.  }
\end{figure*}

At this point it is useful to project out left handed fermion fields 
\begin{equation}
\psi_L={1+\gamma_5\over 2}\psi.
\end{equation}
Then, because of the anomaly, a singlet rotation of only the left
handed field
\begin{equation}
\psi_L\rightarrow e^{i\phi}\psi_L
\end{equation}
is not a good symmetry for generic $\phi$.  On the other hand, a
flavored rotation
\begin{equation}
\psi_L\rightarrow
g_L\psi_L = e^{i\phi_\alpha \lambda_\alpha}\psi_L
\end{equation}
is a symmetry for $g_L\in SU(N_f)$.  The point I wish to emphasize is
that for special discrete elements these two types of rotation can
cross.  In particular I can take $g$ in the center of $SU(N_f)$
\begin{equation}
g=e^{2\pi i/N_f} \in Z_{N_f} \subset SU(N_f),
\end{equation}
and we obtain a valid discrete singlet symmetry
\begin{equation}
\matrix{ &\sigma\rightarrow \sigma \cos(2\pi/N_f)+\eta^\prime
\sin(2\pi/N_f)\cr &\eta^\prime\rightarrow 
\eta^\prime \cos(2\pi/N_f)
-\sigma\sin(2\pi/N_f).
\cr }
\end{equation}
This $Z_{N_f}$ symmetry applies to the effective potential when the
quark mass vanishes.  Then there are $N_f$ equivalent minima in the
$(\sigma,\eta^\prime)$ plane.  This is sketched for the $N_f=4$ case
in Fig.~\ref{potential1}.

At the chiral Lagrangian level this symmetry arises because $Z_N$ is a
subgroup of both $SU(N)$ and $U(1)$.  At the quark level it can be
understood from the fact that the 't Hooft vertex gets a contribution
from each flavor and multiplying together the phases
$\psi_L\rightarrow e^{2\pi i/N_f} \psi_L$ from each gives a net factor
of unity.

\section{Including the quark mass}

A quark mass term $-m\overline\psi\psi\sim -m\sigma$ can be thought of
as tilting the effective potential downward in the sigma direction.
This picks one vacuum as the lowest.  Expanding the potential about
the $n$'th minimum gives an effective pion mass in the given minimum
going as $m_\pi^2 \sim m \cos(2\pi n/N_f)$.  Thus $n=0$ is the true
vacuum while the highest minima are unstable in the $\pi_\alpha$
direction.  Note that multiple truly meta-stable minima become
possible when $N_f>4$.

\begin{figure*}
\centering
\includegraphics[width=2in]{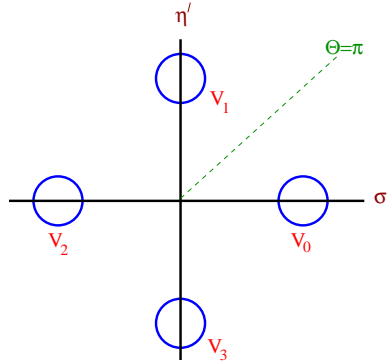}
\caption{\label{potential}
With massive quarks and a twisting angle of $\phi=\pi/N_f$, two of the
minima in the $\sigma,\eta^\prime$ plane become degenerate.  This
corresponds to a first order transition at $\Theta=\pi$.
}
\end{figure*}

While the conventional mass term is proportional to
$m\overline\psi\psi$, it is interesting to consider a more general
term obtained by an anomalous rotation
\begin{equation}
m\overline\psi\psi\rightarrow
m\cos(\phi)\overline\psi\psi
+im\sin(\phi)\overline\psi\gamma_5\psi.
\end{equation}
This corresponds to tilting the potential downward not in the sigma
direction, but in a direction at an the angle $\phi$ in the
$\sigma,\eta^\prime$ plane.  In general this will give an inequivalent
theory.  For small $\phi$, the vacuum will remain in the vicinity of
the minimum at positive $\sigma$; however, as $\phi$ increases through
$\pi/N_f$, vacuum will jump from this minimum to a neighboring one.
This is illustrated in Fig.~\ref{potential}.

In this discussion I have given the mass for each flavor a common phase
$\phi$.  In more conventional treatments one introduces the sum of
these with the definition $\Theta=N_f\phi$.  The $Z_{N_f}$ symmetry
implies a $2\pi$ periodicity in $\Theta$.  What has been demonstrated
here is that with degenerate light quarks a first order transition is
expected at $\Theta=\pi$.

The underlying $Z_{N_f}$ can be thought of as a
discrete symmetry in mass parameter space 
\begin{equation}
m\rightarrow m\exp\left(i\pi\gamma_5\over N_f\right).
\end{equation}
In particular for $N_f=4$ a mass term of form $m\overline\psi \psi$ is
physically equivalent to considering one of form
$im\overline\psi\gamma_5\psi$.  This specific equivalence is only true
for $N_f$ a multiple of 4.

\section{Odd $N_f$}

\begin{figure*}
\centering
\includegraphics[width=2in]{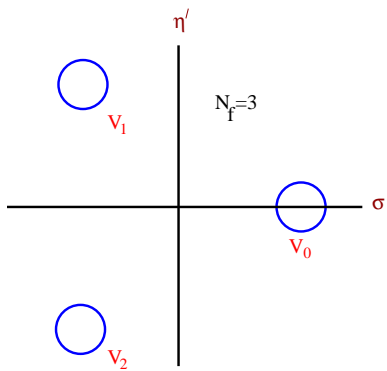}
\caption{\label{potential2}
For odd $N_f$, such as the $SU(3)$ case sketched here, QCD is not
symmetric under changing the sign of the quark mass.
Negative mass corresponds to taking $\Theta=\pi$.
}
\end{figure*}

At this point it should be beginning to be clear why I had restricted
myself to even $N_f$.  Now consider an odd number of flavors,
$N_f=2N+1$.  The crucial point is that $-1$ is not an element of
$SU(2N+1)$.  This means that $m>0$ and $m<0$ not equivalent!  In
particular a negative mass represents $\Theta=\pi$ and will exhibit
spontaneous CP violation with $\langle\eta^\prime\rangle\ne 0$.
Fig.~\ref{potential2} sketches the situation for $SU(3)$.

The fact that the sign of the quark mass is relevant for an odd number
of flavors is something not seen in perturbation theory.  In any given
Feynman diagram, the sign of the mass can be flipped by a $\gamma_5$
rotation.  Thus positive and negative mass three flavor QCD have
identical perturbative expansions and yet are physically different.
This is a simple example of the remarkable fact that inequivalent
theories can have identical perturbative expansions!

\begin{figure*}
\centering
\includegraphics[width=2in]{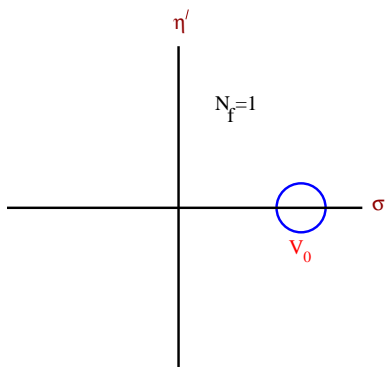}
\caption{\label{potential3} The effective potential for one-flavor QCD
with small quark mass has a unique minimum in the $\sigma,\eta^\prime$
plane.  The minimum is shifted from zero due to the effect of the 't
Hooft vertex.  }
\end{figure*}

A special case of an odd number of flavors is one-flavor QCD.  In this
situation the anomaly removes all chiral symmetry and there is a
unique minimum in the $\sigma,\eta^\prime$ plane, as sketched in
Fig.~\ref{potential3}.  This minimum does not occur at the origin,
being shifted to $\langle \overline\psi \psi\rangle > 0$ by the 't
Hooft vertex, which for one flavor is just an additive mass shift
\cite{Creutz:2007yr}.  Unlike the case with more flavors, this
expectation cannot be regarded as a spontaneous symmetry breaking
since there is no chiral symmetry to break.  Any regulator that
preserves a remnant of chiral symmetry in the one flavor theory must
inevitably fail \cite{Creutz:2008nk}.  Note also that there is no
longer the necessity of a first order phase transition at
$\Theta=\pi$.  It has been argued \cite{Creutz:2006ts} that for finite
quark mass such a transition will occur if the mass is sufficiently
negative, but physics is analytic in $m$ in a finite region around
vanishing mass.

It is important to remember that the details of the instanton effects
are scheme dependent; this is sometimes called the ``renormalon''
ambiguity \cite{Bigi:1994em}.  For the one flavor case this means the
usual polar coordinates $(m,\Theta)$ are singular.  Indeed, it is more
natural to use $({\rm Re}\ m, {\rm Im}\ m)$ as our fundamental
parameters.  The ambiguity is tied to rough gauge configurations of
ill-defined winding number.  Even the overlap operator does not solve
this issue since it is not unique, depending on a parameter often
called the ``domain wall height.''  Because of this, $m=0$ for a
non-degenerate quark is an ambiguous concept.  In Appendix A I discuss
this ambiguity from the point of view of the renormalization group.

\section{When is rooting okay?}

Starting with four flavors, can one adjust $N_f$ down to one using the
formal expression
\begin{equation}
\label{root}
\left| \matrix{
D+m & 0 & 0 & 0 \cr 0 & D+m & 0 & 0 \cr 0 & 0 & D+m & 0 \cr 0 & 0 & 0
& D+m \cr }\right |^{1\over 4} =|D+m|?
\end{equation}
This has been proposed and is widely used as a method for eliminating
the extra species appearing with staggered fermion simulations.

It is important to emphasize that asking about the viability of
Eq.~\ref{root} is a vacuous question outside the context of a
regulator.  Field theory has divergences that need to be controlled,
and, as we have seen above, the appearance of anomalies requires care.
In particular, the regulated theory must break all anomalous
symmetries.

So we must apply Eq.~\ref{root} before removing the regulator.  This
is generally expected to be okay as long as the regulator breaks any
anomalous symmetries appropriately on each of the four factors.  For
example, we expect rooting to be valid for four copies of the overlap
operator.  This satisfies a modified chiral symmetry
$D\gamma_5=-\hat\gamma_5 D$ where the gauge winding $\nu$ appears in
the gauge dependent matrix $\hat\gamma$ through ${\rm
Tr}\hat\gamma_5=2\nu$.

But now suppose we try to force the $Z_4$ symmetry in mass parameter
space before we root.  This is easily done by considering the
determinant
\begin{equation}
\left| \matrix{
D+me^{i\pi\gamma_5\over 4} & 0 & 0 & 0 \cr
0 & D+me^{-i\pi\gamma_5\over 4} & 0 & 0 \cr
0 & 0 & D+me^{3i\pi\gamma_5\over 4} & 0 \cr
0 & 0 & 0 & D+me^{-3i\pi\gamma_5\over 4} \cr
}\right |.
\end{equation}
This maintains the $m\rightarrow me^{i\pi\gamma_5/2}$ symmetry through
a permutation of the four flavors.  This still gives a valid
formulation of the four flavor theory at vanishing $\Theta$ because
the imposed phases cancel.  But expressed in this way, we start with
four one-flavor theories with different values of $\Theta$.  Were we
to root this form, we would be averaging over four inequivalent
theories.  This is not expected to be correct, much as we would not
expect rooting two different masses to give a theory of the average
mass; {\it i.e.}
\begin{equation}
\left(|D+m_1||D+m_2|\right)^{1/2}\ne \left |D+\sqrt{m_1m_2}\right|.
\end{equation}

So we have both a correct and an incorrect way to root a four flavor
theory down to one.  What is the situation with staggered fermions,
the primary place where rooting has been applied?  The problem is that
the kinetic term of the staggered action maintains one exact chiral
symmetry.  Without rooting this is an allowed symmetry amongst what
are usually called ``tastes.''  Under this symmetry there are two
tastes of each chirality.  But, because of this exact symmetry, which
contains a $Z_4$ subgroup, rooting to reduce the theory to one flavor
is not expected to be valid.  In particular, rooting does not remove
the $Z_4$ discrete symmetry in the mass parameter, a symmetry which is
anomalous in the one flavor theory.  Thus, as in the above example,
the tastes are not equivalent and rooting averages inequivalent
theories.

The conclusion is that rooted staggered fermions are not QCD.  So,
what is expected to go wrong?  The unbroken $Z_4$ symmetry will give
rise to extra minima in the effective potential as a function of
$\sigma$ and $\eta^\prime$.  Forcing these minima would most likely
drive the $\eta^\prime$ mass down from its physical value.  This shift
should be rather large, of order $\Lambda_{QCD}$.  This is testable,
but being dominated by disconnected diagrams, may be rather difficult
to verify in practice.

\section{Summary}

We have seen that QCD with $N_f$ massless flavors has a discrete
flavor-singlet $Z_{N_f}$ chiral symmetry.  Associated with this is a
first order transition at $\Theta=\pi$ when $m\ne 0$.  As a
consequence, the sign of the mass is significant for $N_f$ odd, a
property not seen in perturbation theory.  Going down to the $N_f=1$
case, no chiral symmetry survives, leaving $m=0$ unprotected from
additive renormalization.  And finally, this structure is inconsistent
with rooted staggered quarks due to an anomalous $Z_4$ symmetry being
improperly preserved.

\section* {Appendix A: The renormalization group and the quark mass} 

The ambiguity in defining the mass of a non-degenerate quark can be
nicely formulated in the renormalization group framework
\cite{Creutz:2003xc}.  The renormalization group equation for the bare
quark mass
\begin{equation}
a{dm\over da}=m\gamma(g)=m(\gamma_0 g^2+\gamma_1 g^4 +\ldots)
+{\rm non{\hbox{-}}perturbative}
\end{equation}
can in general contain a non-perturbative part that vanishes faster in
$g$ than any power.  From the perturbative part and using the
corresponding flow equation for the bare coupling $g$, we learn that
the bare quark mass runs to zero logarithmically with the cutoff
\begin{equation}
m\propto g^{\gamma_0/\beta_0}
(1+O(g^2))\rightarrow_{a\rightarrow 0} 0.
\end{equation}
where $g$ is the bare gauge coupling, which, by asymptotic freedom,
runs to zero.  We can thus define a renormalized quark mass
\begin{equation}
m_r=\lim_{a\rightarrow 0}\ m g^{-\gamma_0/\beta_0}.
\end{equation}

In general the numerical value of {$m_r$} depends on the details of
the regularization scheme used.  The anomaly, through the 't Hooft
vertex, contributes a non-perturbative part $\sim m^{N_f-1}$ to the
mass flow.  For the case of $N_f=1$, this ceases to vanish in the
massless limit.  Indeed, remembering that
\begin{equation}
m_{\eta^\prime}\propto  {1\over a}
{ e^{-1/2\beta_0 g^2} g^{-\beta_1/\beta_0^2}}
\end{equation}
we might expect a similar form to appear in renormalization group
equation for the mass.  This is particularly so if the $\eta^\prime$
mass is used as a physical observable defining the renormalization
scheme.  Note that this non-perturbative expression formally diverges
if we take $a$ to zero without the appropriate simultaneous decrease
of the coupling.  Allowing such a term can give rise to an additive
shift in the renormalized quark mass.  As an extreme example, consider
a new scheme defined by
\begin{equation}
\matrix{
&\tilde a = a\cr
&\tilde g = g\cr
&\tilde m=m-m_r g^{\gamma_0/\beta_0}\times
{ e^{-1/2\beta_0 g^2} g^{-\beta_1/\beta_0^2}\over \Lambda a}.\cr
}
\end{equation}
This is crafted so that on the renormalization group trajectory the
last factor approaches unity.  With this particular non-perturbative
redefinition of parameters we have
\begin{equation}
\tilde m_r\equiv
\lim_{a\rightarrow 0} \tilde m \tilde g^{-\gamma_0/\beta_0} = 
m_r-m_r=0.
\end{equation}
Thus in the one flavor theory it is always possible to find a scheme
where the renormalized quark mass vanishes!  We conclude that $m=0$
for a non-degenerate quark is an ambiguous concept.  Of course, with
degenerate quarks $m_\pi=0$ defines $m=0$.

\end{document}